\documentclass[10pt,iop]{emulateapj}
\usepackage{url}

\shorttitle{PSA32}
\shortauthors{Jacobs et al.}

\begin{document}

\submitted{Accepted by ApJ Letters: 6 May 2011}
\title{New 145-MHz Source Measurements by PAPER in the Southern Sky}



\author{Daniel C. Jacobs\altaffilmark{1,2}, James E. Aguirre\altaffilmark{1}, Aaron R. Parsons\altaffilmark{3},
 Jonathan C. Pober\altaffilmark{3},  Richard F. Bradley\altaffilmark{4,5,6}, Chris L. Carilli\altaffilmark{7}, 
 Nicole E. Gugliucci\altaffilmark{6}, Jason R. Manley\altaffilmark{8}, Carel van der Merwe\altaffilmark{8}, David F. Moore\altaffilmark{1}, Chaitali R. Parashare\altaffilmark{4}}

\altaffiltext{1}{Dept of Physics and Astron. U. of Pennsylvania, Philadelphia, PA} 
\altaffiltext{2}{Corresponding Author: \email{jacobsda@sas.upenn.edu}}
\altaffiltext{3}{Astron. Dept., U. California, Berkeley, CA} 

\altaffiltext{4}{Dept. of Electrical and Computer Eng., U. Virginia,Charlottesville, VA}
\altaffiltext{5}{Natl. Radio Astron. Obs., Charlottesville, VA}
\altaffiltext{6}{Astronomy Dept., U. Virginia, Charlottesville, VA}

\altaffiltext{7}{Natl. Radio Astron. Obs., Socorro, NM}
\altaffiltext{8}{Karoo Array Telescope, Capetown, South Africa}

\begin{abstract}
We present observations from the Precision Array for Probing the
Epoch of Reionization (PAPER) in South Africa, observed in May and
September 2010. Using two nights of drift scanning with PAPER's
60\arcdeg\ FWHM beam we have made a map covering the entire sky below
+10 degrees declination with an effective center frequency of 145 MHz,
a 70-MHz bandwidth, and a resolution of 26\arcmin.  A 4800 square-degree
region of this large map with the lowest Galactic emission
reaches an RMS of 0.7 Jy. We establish an absolute flux scale using
sources from the 160-MHz Culgoora catalog.  Using the 408-MHz Molonglo
Reference Catalog (MRC) as a finding survey, we identify counterparts
to 480 sources in our maps, and compare our fluxes to the MRC and to
332 sources in the Culgoora catalog.  For both catalogs, the ratio of
PAPER to catalog flux averages to 1, with a standard deviation of 50\%.
This measured variation is consistent with comparisons between
independent catalogs observed at different bands.  The PAPER data represent new
145-MHz flux measurements for a large number of sources in the band expected
to encompass cosmic reionization, and represents a significant step 
toward establishing a model for removing foregrounds to the reionization
signal.

\end{abstract}


\section{Introduction}

Emission from the highly redshifted 21 cm line of neutral hydrogen is
a very promising method of exploring the Epoch of Reionization (EoR)
over the redshift range $6 < z < 12$ ({Furlanetto}, {Oh}, \&  {Briggs} 2006; {Fan}, {Carilli}, \& {Keating} 2006).  Several
EoR experiments are currently operating, including the Murchison Widefield Array (MWA;
{Lonsdale} {et~al.} 2009), the Low Frequency Array (LOFAR; {R\"{o}ttgering} {et~al.} 2006)   the Giant Metre-wave Radio Telescope (GMRT; {Paciga} {et~al.} 2011), and 
the Precision Array for Probing the Epoch of Reionization (PAPER;
{Parsons} {et~al.} 2010).  Both PAPER and the MWA operate in the southern
hemisphere, as will the future Square Kilometer Array (SKA).  It is anticipated that
to effectively suppress foreground emission many orders of
magnitude brighter than the EoR signal,
these instruments will need
high levels of observational stability and exquisite characterization
and control of instrumental systematics.  
These goals will be greatly
aided by building a complete sky model that includes accurate 
point-source locations and fluxes.

Several recent attempts have been made to synthesize what is known
about the radio sky from existing surveys.
{de Oliveira-Costa} {et~al.} (2008) compiles measurements from 10 MHz to 90 GHz
and uses them to create a global sky model of extended emission (sizes greater than two degrees). Discrete sources
are listed in many catalogs. Those used for comparison here are listed in  Table \ref{tab:surveys}.

The literature grows increasingly sparse at lower
frequencies and in the southern hemisphere; indeed, {\it no} blind
survey has been reported below 408 MHz in the south.  The 
SPECFIND cross-identification catalog identifies 6000 unique sources between 100 and 200 MHz
with  $\delta>0\arcdeg$, but fewer than 1000  sources
with declination $\delta<0\arcdeg$.  The best information in the south 
($\delta<-30\arcdeg$) below 408 MHz is provided by the Molonglo 4 Jy Survey
(MS4; {Burgess} \& {Hunstead} 2006), which uses the MRC (complete to 1 Jy),
Culgoora observations by {Slee} (1995) at 160 MHz and
80 MHz, and new Molonglo observations at 408 and 800 MHz to estimate
the 178-MHz flux of bright sources in a sample similar to
the northern 3CRR survey ({Laing}, {Riley}, \& {Longair} 1983).

Catalog comparisons are hindered both by spectral/temporal variation of
sources and by differences in the angular resolution of observations.  Point-source confusion is of particular
concern, given the large synthesized beams of many low-frequency instruments.
Spectra at low frequencies tend to be dominated by synchrotron emission.
While this emission tends to be well-described by a power-law in frequency,
at lower frequencies,
several effects complicate comparisons
between bands.  Chief
among these is synchrotron self-absorption, which is most often in
evidence at the lowest frequencies,
but may also produce spectral
features between 100 and 200 MHz ({Helmboldt} {et~al.} 2008).  Another complication arises from sources not present in
the deep high-frequency catalogs appearing at lower frequencies due
to exceptionally steep spectral indices.  
A source steep enough to appear in these first PAPER images ($>10Jy$) and be absent 
in the MRC sample ($>1Jy$) would need a very steep spectral index  $<-2.2$.
Out of over 70000
sources in NVSS only two or three sources are known to have spectral
indices this steep ({van Weeren} {et~al.} 2009).  For these reasons, it is
desirable to have measurements in the band of interest at an
appropriate resolution.

In this Letter, we present new flux
measurements of 480 sources at 145 MHz using PAPER.
These measurements cover the largest
area of the southern sky yet surveyed in the EoR band.
We describe PAPER and the observations in 
\S\ref{sec:Observations}, detail our method for calibration and
counterpart identification in \S\ref{sec:Catalog} and conclude
in \S\ref{sec:Conclusion} with prospects for future improvements.

\section{Observations and Data Reduction}
\label{sec:Observations}

PAPER is a first-generation experiment
focusing on the statistical detection of the fluctuations in HI
emission during the EoR.  It is an interferometric transit array
operating between 110 and 180 MHz ($7 < z < 12$ for HI).
PAPER consists of two deployments: a 32-antenna array at the
NRAO facilities near Green Bank, WV, and a 32-antenna array
in South Africa's Radio Quiet Zone (PAPER South Africa, hereafter PSA32).  
The final PSA instrument will consist of 128 full-Stokes dipole antennas.  The data
presented here are from the initial deployment of PSA32.

Each PAPER antenna is a pair of crossed sleeve dipoles mounted above a
mesh ground screen.
Signals from each antenna are amplified, directly
digitized, and then correlated by an FPGA-based correlator ({Parsons} 2008).
PAPER's primary beam is 60\arcdeg\ FWHM with a spatially-
and spectrally-smooth
response nearly to the horizon.  A point at zenith takes about
4 hours to cross PAPER's primary beam, during which time apparent flux
changes by a factor of two. 

The PSA32 antennas are arranged in a minimum redundancy configuration,
constructed to provide uniform sampling of the $uv$-plane within a 300-meter-wide
circle while avoiding a central radius of 10 m ({Parsons} {et~al.} 2011).
The maximum baseline length of this arrangement generates an
effective resolution of 26\arcmin\ at 145 MHz.  Antenna positions
were surveyed using a commercial differential GPS to centimeter precision.  We found
no further position refinement necessary.


During May and September 2010, we recorded commissioning data
with PSA32 in two separate campaigns.  Between the May and September
data-taking, a number of small improvements to the correlator were
made, but all other hardware remained unchanged.  The data presented
here are from UT 2010 May 19 13:11 - 2010 May 20 04:50 (15 hours) and
UT 2010 Sep 15 16:48 - 2010 Sep 16 04:04 (12 hours), both periods
being predominantly between sunset and sunrise.  Only the linear EW dipoles of each
antenna were correlated.  Visibilities were integrated and recorded
every 5.37 seconds.  The frequency resolution was 96 kHz in May and 45
kHz in September.  The separation in LST between the two observing
epochs, along with PAPER's wide primary beam, make these two observations 
sufficient to map the entire sky below $\delta<10\arcdeg$.

Data editing, calibration and imaging were performed using 
the Astronomical Interferometry in PYthon (AIPY)
package\footnote{\url{http://casper.berkeley.edu/astrobaki/index.php/AIPY}}.
The first analysis step was to obtain a phase calibration.  Because of the wide
field-of-view (FOV) of the antennas, the data are dominated by bright
sources that are sometimes far from the zenith.
During the May observation, the brightest source visible was Cen
A\footnote{While Cen A is resolved, the central point source
dominates the smooth structure by several orders of magnitude.}, while
the brightest source during September observations was Pic A. These sources are bright
enough to perform single-baseline fringe fitting.  Data observed within ten minutes
of the transit of these sources were used to derive a phase
calibration by fitting a time and frequency visibility model to the
data using a conjugate-gradient solver ({Parsons} {et~al.} 2010).  Phase terms in the
calibration are dominated by cable and correlator delays; these have
been found to be quite stable.  Thus for the analysis here, the phase
calibration derived from these two 10-minute observations is applied to each night's entire observation.

If the source is not carefully removed, the phase-calibrator sidelobes severely
limit imaging dynamic range.  An efficient source-removal technique 
filters data by removing the corresponding region of
delay/delay-rate (DDR) space for each baseline ({Parsons} \& {Backer} 2009).  This
technique
filters a source from each baseline by nulling data
having a delay and fringe-rate corresponding to the desired sky
location.  

The net effect is
to remove a large fraction of the filtered source {\it without} having
to construct a multi-source image-domain model. For the May data, the point-source
component of Cen A (estimated flux $\sim5000$~Jy) is 
filtered; for September, we have removed Pic A and For A 
(400 and 150 Jys, respectively).

Given the high quality of the RFI environment at the Karoo site, RFI-flagging was limited 
to flagging of a few satellite bands, as well
as any visibilities with amplitudes $2\sigma$ above the mean amplitude as in {Parsons} {et~al.} (2010). Less than one percent of the data
are flagged in this way.

In this quiet environment, instrumental effects became dominant.
A troublesome instrumental effect in many interferometric instruments is common-mode interference, interfering signals common to two or more inputs and 
sky signals crossing antenna boundaries within the analog
system. Both of these kinds of cross-talk  are  removed by subtracting a 4 hour long running average 
as described in {Parsons} {et~al.} (2010).

Map-making is done in three stages: snapshot-imaging,
mosaicking each night and finally summing into a single calibrated
map.

Images are made in 10-minute zenith-phased ``snapshots''; this is a
sufficiently short time that the effects sources moving through the
primary beam is negligible.  Visibility data are gridded into the
$uv$-plane using linear multi-frequency synthesis ({Taylor}, {Carilli}, \& {Perley} 1999) and $w$-projection ({Cornwell}, {Golap}, \&  {Bhatnagar} 2008).
To this $uv$-gridded data we apply radial weighting ---increasing radially in the $uv$-plane--- to emphasize point sources.  Gridded data are Fourier transformed to produce a snapshot image 70\arcdeg\ wide, with an effective synthesized beam width of
26\arcmin.  These facets are then deconvolved by the equivalent dirty
beam using
the H{\"o}gbom CLEAN algorithm ({H{\"o}gbom} 1974).  Image-domain deconvolution
is limited in its ability to reconstruct the flux, particularly in the
wide-field case ({Rau} {et~al.} 2009).  Thus, the CLEANing is not fully
effective, and the images contain artifacts from the 
side-lobes of sources far from the phase center.

All snapshot images made over the course of a night are weighted
by a model of the primary
beam and averaged onto a HEALPIX grid ({G{\'o}rski} {et~al.} 2005) with 7\arcmin\
pixels (NSIDE=512), to create two maps --- one for each epoch. A typical
pixel has weighted contributions from approximately five snapshots.
Each map is flux-calibrated to a bright source near the
phase-calibrator using a flux taken from the Culgoora catalog. The May
map is flux-calibrated to 1422-297 at 21 Jy and the September map is
calibrated to 0521-365 at 72 Jy. Once on the same flux-scale, the two epoch
maps are summed together into a single map, weighted by the number of
snapshot contributions. The final product covers 36000 square degrees
at $\delta<10\arcdeg$, with an effective resolution of
26\arcmin. 

The limitations of these reduction steps, as well as instrumental artifacts,
impact image fidelity.
Final images include residual cross-talk and errors due
to delay-filtering, which necessarily
removes flux from multiple points on the map. The absence of time-dependent
calibration, the limitation to image-plane
deconvolution and uncertainty in the beam model also affect the accuracy of the map.

Successful future 
work in foreground mapping and EoR detection will depend on our 
ability to correctly identify the most important of these issues. This is true not only
within the PAPER project, but also between similar projects. For these reasons
we establish an accuracy baseline by measuring and comparing the fluxes
of many sources to catalog values.

\section{Catalog Construction and Flux Calibration}
\label{sec:Catalog}

\begin{figure*}
\includegraphics[width=1\textwidth]{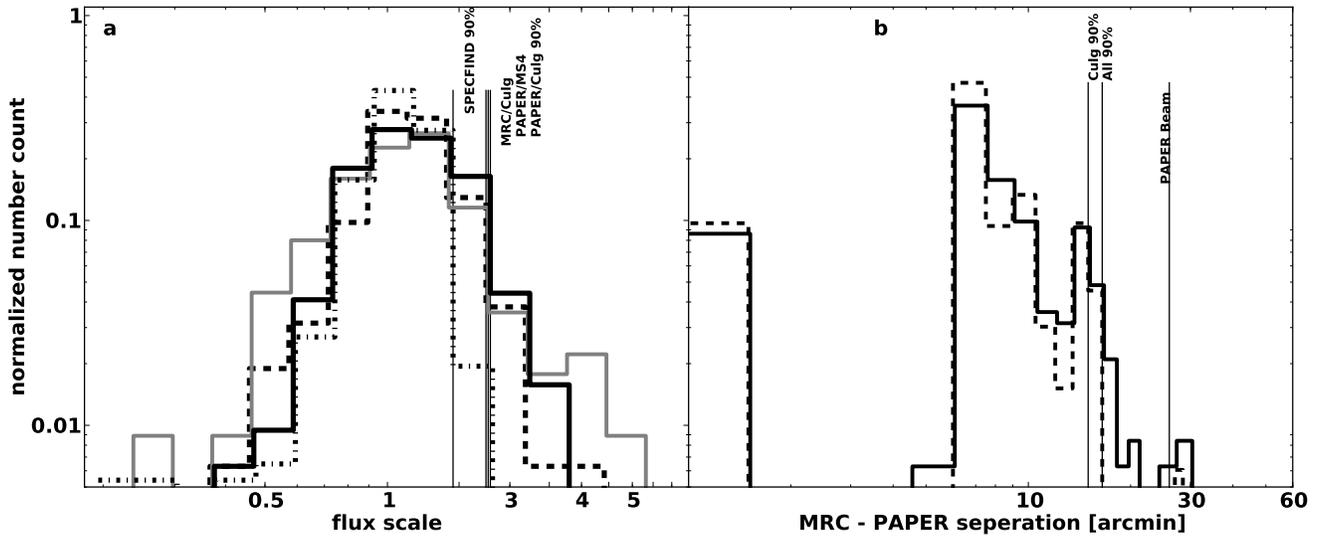}
\caption{
a) PAPER's flux scale ---the ratio of PAPER fluxes to Culgoora fluxes at 160MHz (black, solid)
and to MS4 fluxes interpolated to 178MHz (grey,
solid).  PAPER/Culgoora and PAPER/MS4  with FWHM of 1 and 0.7 have distributions similar to the
MRC/Culgoora flux ratio (thick, dashed) with a FWHMs of 0.5.
A similar comparison between 178 and 151 MHz objects
co-identified by SPECFIND (dot-dashed) shows that a somewhat tighter
agreement (0.5 FWHM with fewer large outliers) is possible if cross-matching
is done while accounting for morphology and instrumentation. b) Distance in degrees between MRC position and identified PAPER peak. 
All MRC sources within $-85\arcdeg<\delta<10\arcdeg$ have been paired with a PAPER peak within the plotted range.
PAPER positions are defined by the centers of HEALpix pixels that are ~$7'$ on a side leading to 
quantization effects; 90\% of sources are within one PAPER beam (thin solid vertical line).}
\label{fig:fluxscale}\label{fig:seps}
\end{figure*}

We have used the entire sky below $\delta<10\arcdeg$ to find fluxes corresponding to 480 MRC sources above 4
Jy  --- selection criteria similar to those used by {Burgess} \& {Hunstead} (2006) to
generate the MS4 sample.  The PAPER flux is identified as the 
brightest pixel within $30\arcmin$ of the MRC source.  They are listed along with
separation distances and local RMS in Table \ref{tab:catalog};  90\% of the sources identified are within one beam-width (see Figure \ref{fig:seps}). In the following we will explore the relative completeness and accuracy of this catalog.

\begin{figure*}
\vspace{1in}
\includegraphics[width=1\textwidth]{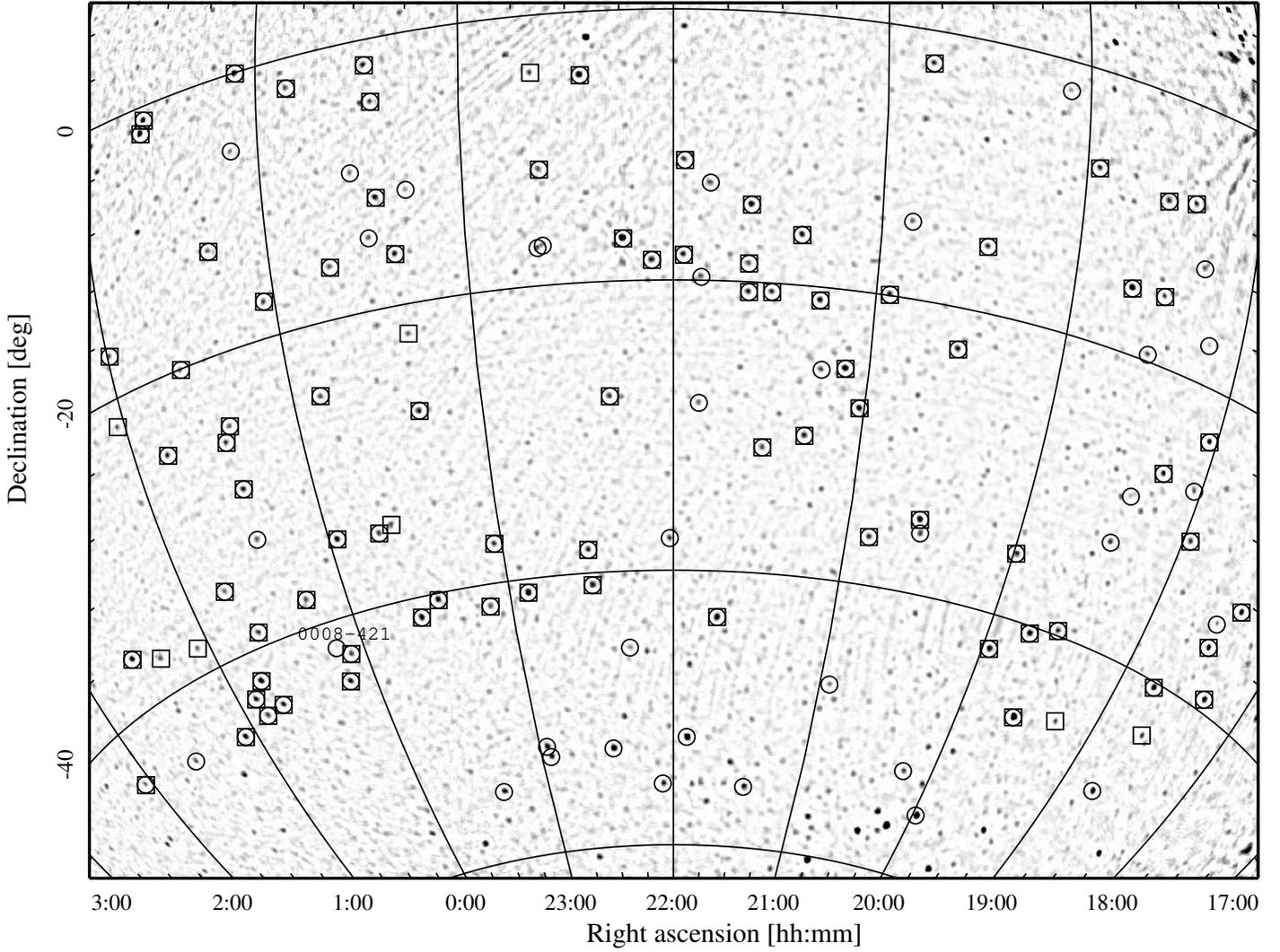}\caption{A low foreground region of the PSA32 field, centered at $23^\mathrm{h}$ $-31\arcdeg$.
Pixels above 99\% of the flux scale, approximately 1 Jy, are shown
in black. MRC sources above 4 Jy are shown in circles. PAPER fluxes for these sources are given in Table \ref{tab:catalog}. All MRC markers, save one, have a corresponding source
at this flux level. The 160MHz Culgoora survey, used to evaluate the
flux scale, is shown in squares. This image has an effective integration
time of 30 minutes, a bandwidth of 70MHz, a field rms of 0.4 Jy, and a peak to field dynamic range of 120.}
\label{fig:image}
\end{figure*}

\begin{figure*}
\includegraphics[width=1\textwidth]{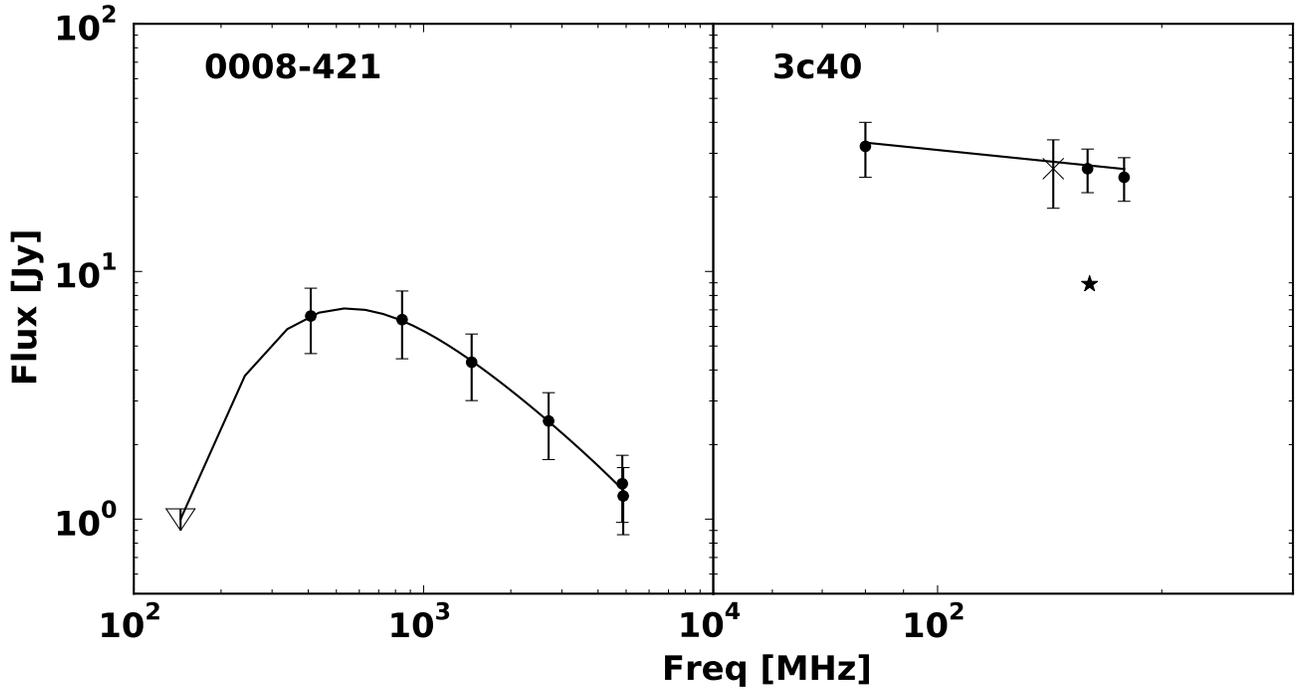}

\caption{Left: Spectrum of 0008-421, the only MRC source in Figure \ref{fig:image} without a PAPER counterpart. A rare example of spectral turnover at around 500 MHz.  Right: Spectrum of 3c40 (0123-016) with an 'x' for PAPER flux and star for Culgoora 160-MHz. All circles taken from SPECFIND.
}
\label{fig:spectra}

\end{figure*}

The accuracy of the PAPER image and of these fluxes can be evaluated both superficially in the image plane and numerically by comparing to past measurements. By comparing the MRC catalog directly to the image we can ascertain the relative completeness of the MRC sample. In Figure \ref{fig:image} we overlay MRC markers from our 4-Jy subsample onto a 4800 square-degree sub-image and see that at the 4-Jy flux-level MRC is not one-to-one but does agree with the map on all but one source. As shown in Figure \ref{fig:spectra} this source manifests a rare high-frequency turn-over.

In the case of the Culgoora catalog, there are no
such differences. In places where Culgoora shows a
bright source that is not in the 408 MHz sample, PAPER also finds a bright source. Sources as dim as $~10\mathrm{Jy}$ do not have a matching MRC marker. Together these facts suggest that by limiting to MRC sources above 4 Jy we have excluded sources with steep ($\alpha<-1$, where $S=S_0 (\frac{\nu}{\nu_0})^\alpha$) spectra. These steeper sources are below our flux limit at 408MHz but are bright at 145MHz. Thus our complete flux-limited sample of MRC at 408MHz becomes incomplete at 145MHz.  

To assess the accuracy of the flux measurements we compare with 332 sources also found in the Culgoora catalog and 225 found in MS4. 
The accuracy of the PAPER measurements will be limited by the image dynamic range, as discussed above, as well as the relatively wide bandwidth of the PAPER.  However these errors must be set against the error in the catalog comparison. As can be seen in the Helmboldt or SPECFIND collections of radio spectra, the presence of multiple components or extended
structure in sources hampers comparisons between observations with different resolutions, while the presence of
self-absorption or other spectral structure impedes comparison between catalogs generated at 
different frequencies. To set the scale of these effects we intercompare several catalogs with measurements near the PAPER band. 

A simple comparison metric is the per-source flux-scale; the ratio of fluxes between two sets.  Accounting for spectral slope, the flux-scale would have a nominal value of one, with a certain amount spread encompassing all the sources of error in flux determination and catalog comparison. In Figure \ref{fig:fluxscale} we have plotted the distributions for the PAPER/Culgoora and PAPER/MRC\footnote{Where
fluxes were measured at frequencies outside of the PAPER band (110-180
MHz), we have scaled the flux using a spectral index of $\alpha=-1$
(the average index at these frequencies;
{Slee} 1995; {Helmboldt} {et~al.} 2008; {Bennett} 1962) } To estimate the catalog comparison error we have calculated the MRC/Culgoora flux-scale, as well as between all sources in SPECFIND at 151-MHz and those at 178-MHz. The SPECFIND comparison has the
advantage of more accurate
cross-identification between sources. In addition, most of the measurements in these bands were done by  the Cambridge
Low Frequency Telescope (6 and 7C) and the 3CR all of which are known
to be in good agreement ({Bennett} 1962; {Gower}, {Scott}, \& {Wills} 1967; {Baldwin} {et~al.} 1985).  
Even so, a noticeable (although narrower) spread of flux scales is
apparent (Figure \ref{fig:fluxscale}) In the SPECFIND comparison 90\% of sources are below a flux-scale of 1.5, while in all other comparisons the 90\% level occurs at 1.75.  The distribution of the MRC/Culgoora flux comparison has a spread similar
to the distribution of PAPER's fluxes relative to each of these catalogs. 

The MRC sources have been cross-identified with the Culgoora using the same algorithm as the MRC-PAPER cross-identification. Thus the MRC-Culgoora comparison would be more likely to have similar catalog comparison errors and in fact does have a similar distribution of flux-scales. This similarity implies that systematic errors will not be easily distinguished by catalog comparison. As an example consider the most extreme flux-scale outlier 0123-016 (3c40). PAPER observes a flux of 26 Jy while Culgoora only 8.9 Jy.  Culgoora notes this source to have multiple components and measures 32 Jy in the 80-MHz band. When we add 3C and 3CR to the spectrum we see a consistent picture of a source around 30 Jy as shown in Figure \ref{fig:spectra} with the 160-MHz point the only in disagreement. Catalog discrepancies of \emph{this} type are rare but there are many types.


\section{Discussion and Conclusions}
\label{sec:Conclusion}

The Epoch of Reionization signal will be faint; detection will require
precise calibration as well as deep foreground removal. Self-calibration
of a wide-field instrument requires knowledge of sources
covering a large fraction of the sky.  This calibration
may then be verified by comparing multiple measurements,
ideally between telescopes.

To facilitate such comparison, we publish here the first
catalog derived from early PAPER data which, despite suffering from various 
systematics related to instrument response and analysis methodology,
shows qualitatively good agreement with other measurements.
Our demonstrated ability to map more than half of the sky with two days of observation represents
a major advance in survey speed and is made possible by the width of
PAPER's primary beam, the bandwidth of PAPER's correlator, and
the use of w-projection.

A number of improvements to the instrument and data processing methodology are
currently underway.  Cross-talk can be mitigated by the addition of one-way RF coupling and
Walsh switching.  Both are likely to reduce excess correlations though their effectiveness are still being evaluated.  

Preliminary work is also under-way to refine imaging using the
Common Astronomy Software Applications (CASA)
environment\footnote{\url{http://casa.NRAO.edu/}}. Early tests of Cotton
CLEAN, faceting combined with w-projection, time-dependent calibration,
and spectroscopic imaging have been favorable; this system will be
used to produce higher dynamic range maps suitable for a blind survey
and spectroscopic exploration. Finally, the dynamic range is limited

by the instantaneous $uv$-coverage of the 32-element antenna configuration.
Future deployments with larger number of antennas 
will result in additional improvements to the snapshot $uv$-coverage and imaging dynamic
range.

Implementing these instrumentation and processing improvements will
help produce images of even higher fidelity that might
reach the sub-Jy confusion limit.  From such images, it will be possible to construct a
complete blind catalog using source fitting and
photometric analysis.  This more precise catalog will merit a
stricter comparison with previous catalogs that more carefully accounts for extended
structure and confusion.

Compared to the current array, the final PAPER South Africa array will
have four times as many elements and should have 16 times the dynamic range.  Here we
have imaged and cataloged what is essentially a dirty image of the sky
and already found good agreement. The final PAPER telescope will be
capable of spectral imaging the 110 to 180-MHz night sky to the confusion limit once a
day. Although the radio sky was first explored at meter wavelengths,
much remains unknown about the spectral and temporal behaviour of
sources in this frequency band. These early PAPER 32-element
commissioning data are already demonstrating a reliable level of
accuracy that are limited primarily by well-known problems.  We can
reasonably expect future PAPER data to add substantially to our
understanding of the sky at meter wavelengths.
\acknowledgements
This work made use of the Topcat 
package\footnote{\url{http://www.starlink.ac.uk/topcat/}} as well as the Vizier 
Virtual Observatory database\footnote{\url{http://vizier.u-strasbg.fr/viz-bin/VizieR}}.
AP would like to acknowledge support through the NSF AAPF program (\#0901961) and
from the Charles H. Townes Postdoctoral Fellowship.  The PAPER project is supported
through the NSF-AST program (\#0804508).


\clearpage 



\begin{deluxetable}{rrrrrrr} 
\tablecolumns{6} 

\tablecaption{Low frequency surveys used in this Letter} 

\tablehead{ 
\colhead{Name} & \colhead{Res} & 
\colhead{Freq [Mhz]} & \colhead{Dec Limits} &
\colhead{\parbox{4em}{Flux Limit [Jy]}} &  \colhead{Ref} }
\startdata 
MRC&	2'&408&$18.5>d>-85$&1& {Large} {et~al.} (1981)\tabularnewline
MS4&	2'&178$^b$,408&$-30>d>-85$ & 4 &  {Burgess} \& {Hunstead} (2006)\tabularnewline
Culgoora&	1.85',3.7'&160,80&$32>d>-50$&2& {Slee} (1995)\tabularnewline
	3CR(R)$^a$ &	6'&178&$75>d>-50$&5&{Bennett} (1962)({Laing} {et~al.} 1983)\tabularnewline
	6C$^a$&	4.2'&151&$>30$&0.3&{Baldwin} {et~al.} (1985)\tabularnewline
	7C$^a$&	1.17'&151&$>26$&0.2&{Gower} {et~al.} (1967)\tabularnewline
VLSS$^a$&80"&	74&$>-30$&0.4&{Cohen} {et~al.} (2007)\tabularnewline
NVSS$^a$&45"&	1400&$>-40$&2.5e-3& {Condon} {et~al.} (1998)\tabularnewline
PAPER&26'&	145&$10>$&10& This Letter\tabularnewline
\enddata
\tablenotetext{a}{ Sources included via the SPECFIND meta-catalog ({Vollmer} {et~al.} 2010) .}
\tablenotetext{b}{178MHz fluxes in MS4 are estimates based on MRC, Culgoora and other measurements.}
\label{tab:surveys}
\end{deluxetable}

\begin{deluxetable}{ccccccccrrr}
\tablecolumns{12}
\tablecaption{PAPER fluxes for 480 MRC sources and matching Culgoora fluxes (where available)}
\tablehead{
\colhead{Ra} & \colhead{Dec} & \colhead{Name} & \colhead{S145} & 
\colhead{rms}  & \colhead{MRC\_sep} & \colhead{Cul} & 
\colhead{S(160)} & \colhead{S(80)} & \colhead{SpIndex}}
\startdata
  0.60 & -83.14 & 0003-833 & 18.9 & 4.7 & 0.17 &  &  &  & \\
  0.88 & -17.50 & 0000-177 & 11.6 & 1.2 & 0.11 & 0000-177 & 11.8 & 22 & -0.9\\
  1.37 & -56.54 & 0003-567 & 12.0 & 2.2 & 0.17 &  &  &  & \\
  1.52 & -42.61 & 0003-428 & 9.6 & 1.3 & 0.16 & 0003-428 & 11.9 & 11 & 0.11\\
  1.58 & -0.07 & 0003-003 & 25.5 & 2.6 & 0.12 & 0003-003 & 16.8 & 27 & -0.68\\
  2.11 & -6.05 & 0005-062 & 11.2 & 1.8 & 0.23 & 0005-062 & 6.9 & 10 & -0.54\\
  2.51 & -44.50 & 0007-446 & 14.3 & 1.2 & 0.16 & 0007-446 & 17.0 & 26 & -0.61\\
  3.27 & -42.11 & 0008-421 & 1.5 & 1.4 & 0.41 &  &  &  & \\

\enddata
\tablecomments{PAPER Southern Sky catalog generated by searching for sources in the
Molonglo Reference Catalog above 4 Jy and below $+10\arcdeg$ Declination. Beginning
on the left, the columns list: Right Ascension and Declination in
degrees, MRC name, calibrated PAPER flux [Jy], rms around source [Jy] and angular separation in degrees from MRC location.  Included
for reference are Culgoora 160MHz, 80MHz and spectral indices fluxes
where available. Complete table available in the online edition these Letters.  }
\label{tab:catalog}
\end{deluxetable}


\begin{thebibliography}{26}
\expandafter\ifx\csname natexlab\endcsname\relax\def\natexlab#1{#1}\fi

\bibitem[{{Baldwin} {et~al.}(1985){Baldwin}, {Boysen}, {Hales}, {Jennings},
  {Waggett}, {Warner}, \& {Wilson}}]{baldwin85}
{Baldwin}, J.~E., {Boysen}, R.~C., {Hales}, S.~E.~G., {Jennings}, J.~E.,
  {Waggett}, P.~C., {Warner}, P.~J., \& {Wilson}, D.~M.~A. 1985, \mnras, 217,
  717

\bibitem[{{Bennett}(1962)}]{bennett62}
{Bennett}, A.~S. 1962, \mnras, 125, 75

\bibitem[{{Burgess} \& {Hunstead}(2006)}]{burgess06}
{Burgess}, A.~M. \& {Hunstead}, R.~W. 2006, \aj, 131, 100

\bibitem[{{Cohen} {et~al.}(2007){Cohen}, {Lane}, {Cotton}, {Kassim}, {Lazio},
  {Perley}, {Condon}, \& {Erickson}}]{cohen07}
{Cohen}, A.~S., {Lane}, W.~M., {Cotton}, W.~D., {Kassim}, N.~E., {Lazio},
  T.~J.~W., {Perley}, R.~A., {Condon}, J.~J., \& {Erickson}, W.~C. 2007, \aj,
  134, 1245

\bibitem[{{Condon} {et~al.}(1998){Condon}, {Cotton}, {Greisen}, {Yin},
  {Perley}, {Taylor}, \& {Broderick}}]{condon98}
{Condon}, J.~J., {Cotton}, W.~D., {Greisen}, E.~W., {Yin}, Q.~F., {Perley},
  R.~A., {Taylor}, G.~B., \& {Broderick}, J.~J. 1998, \aj, 115, 1693

\bibitem[{{Cornwell} {et~al.}(2008){Cornwell}, {Golap}, \&
  {Bhatnagar}}]{cornwell08}
{Cornwell}, T.~J., {Golap}, K., \& {Bhatnagar}, S. 2008, IEEE Journal of
  Selected Topics in Signal Processing, 2, 647

\bibitem[{{de Oliveira-Costa} {et~al.}(2008){de Oliveira-Costa}, {Tegmark},
  {Gaensler}, {Jonas}, {Landecker}, \& {Reich}}]{deoliveira-costa08}
{de Oliveira-Costa}, A., {Tegmark}, M., {Gaensler}, B.~M., {Jonas}, J.,
  {Landecker}, T.~L., \& {Reich}, P. 2008, \mnras, 388, 247

\bibitem[{{Fan} {et~al.}(2006){Fan}, {Carilli}, \& {Keating}}]{Fan06}
{Fan}, X., {Carilli}, C.~L., \& {Keating}, B. 2006, \araa, 44, 415

\bibitem[{{Furlanetto} {et~al.}(2006){Furlanetto}, {Oh}, \&
  {Briggs}}]{furlanetto06}
{Furlanetto}, S.~R., {Oh}, S.~P., \& {Briggs}, F.~H. 2006, \physrep, 433, 181

\bibitem[{{G{\'o}rski} {et~al.}(2005){G{\'o}rski}, {Hivon}, {Banday},
  {Wandelt}, {Hansen}, {Reinecke}, \& {Bartelmann}}]{gorski05}
{G{\'o}rski}, K.~M., {Hivon}, E., {Banday}, A.~J., {Wandelt}, B.~D., {Hansen},
  F.~K., {Reinecke}, M., \& {Bartelmann}, M. 2005, \apj, 622, 759

\bibitem[{{Gower} {et~al.}(1967){Gower}, {Scott}, \& {Wills}}]{gower67}
{Gower}, J.~F.~R., {Scott}, P.~F., \& {Wills}, D. 1967, \memras, 71, 49

\bibitem[{{Helmboldt} {et~al.}(2008){Helmboldt}, {Kassim}, {Cohen}, {Lane}, \&
  {Lazio}}]{helmboldt08}
{Helmboldt}, J.~F., {Kassim}, N.~E., {Cohen}, A.~S., {Lane}, W.~M., \& {Lazio},
  T.~J. 2008, \apjs, 174, 313

\bibitem[{{H{\"o}gbom}(1974)}]{hogbom74}
{H{\"o}gbom}, J.~A. 1974, \aaps, 15, 417

\bibitem[{{Laing} {et~al.}(1983){Laing}, {Riley}, \& {Longair}}]{laing83}
{Laing}, R.~A., {Riley}, J.~M., \& {Longair}, M.~S. 1983, \mnras, 204, 151

\bibitem[{{Large} {et~al.}(1981){Large}, {Mills}, {Little}, {Crawford}, \&
  {Sutton}}]{large81}
{Large}, M.~I., {Mills}, B.~Y., {Little}, A.~G., {Crawford}, D.~F., \&
  {Sutton}, J.~M. 1981, \mnras, 194, 693

\bibitem[{{Lonsdale} {et~al.}(2009){Lonsdale}, {Cappallo}, {Morales}, {Briggs},
  {Benkevitch}, {Bowman}, {Bunton}, {Burns}, {Corey}, {Desouza}, {Doeleman},
  {Derome}, {Deshpande}, {Gopala}, {Greenhill}, {Herne}, {Hewitt}, {Kamini},
  {Kasper}, {Kincaid}, {Kocz}, {Kowald}, {Kratzenberg}, {Kumar}, {Lynch},
  {Madhavi}, {Matejek}, {Mitchell}, {Morgan}, {Oberoi}, {Ord},
  {Pathikulangara}, {Prabu}, {Rogers}, {Roshi}, {Salah}, {Sault}, {Shankar},
  {Srivani}, {Stevens}, {Tingay}, {Vaccarella}, {Waterson}, {Wayth}, {Webster},
  {Whitney}, {Williams}, \& {Williams}}]{lonsdale09}
{Lonsdale}, C.~J. {et~al.} 2009, IEEE Proceedings, 97, 1497

\bibitem[{{Paciga} {et~al.}(2011){Paciga}, {Chang}, {Gupta}, {Nityanada},
  {Odegova}, {Pen}, {Peterson}, {Roy}, \& {Sigurdson}}]{paciga11}
{Paciga}, G. {et~al.} 2011, \mnras, 244

\bibitem[{{Parsons}(2008)}]{parsons08}
{Parsons}, A.~e. 2008, \pasp, 120, 1207

\bibitem[{{Parsons} \& {Backer}(2009)}]{parsons09}
{Parsons}, A.~R. \& {Backer}, D.~C. 2009, \aj, 138, 219

\bibitem[{{Parsons} {et~al.}(2010){Parsons}, {Backer}, {Foster}, {Wright},
  {Bradley}, {Gugliucci}, {Parashare}, {Benoit}, {Aguirre}, {Jacobs},
  {Carilli}, {Herne}, {Lynch}, {Manley}, \& {Werthimer}}]{parsons10}
{Parsons}, A.~R. {et~al.} 2010, \aj, 139, 1468

\bibitem[{{Parsons} {et~al.}(2011){Parsons}, {McQuinn}, {Aguirre}, \&
  {Pober}}]{parsons11}
{Parsons}, A.~R., {McQuinn}, M.~and{Jacobs}, D.~C., {Aguirre}, J., \& {Pober},
  J. 2011, \apj,submitted, arxiv:astro-ph/1103.2135

\bibitem[{{Rau} {et~al.}(2009){Rau}, {Bhatnagar}, {Voronkov}, \&
  {Cornwell}}]{rau09}
{Rau}, U., {Bhatnagar}, S., {Voronkov}, M.~A., \& {Cornwell}, T.~J. 2009, IEEE
  Proceedings, 97, 1472
  
\bibitem[{{R\"{o}ttgering}{et~al.}}(2006)]{rottgering06}
{R\"{o}ttgering}, HJA., {Braun}, R., {Barthel}, PD., {van Haarlem}, MP., {Miley}, GK., {et~al}. 2006, In \emph{Proc. Cosmol., Galaxy Form. Astropart. Phys. Pathw.} SKA, Oxford, April 10Ð12, 2006. Published by ASTRON (astro-ph/0610596) 


\bibitem[{{Slee}(1995)}]{slee95}
{Slee}, O.~B. 1995, Australian Journal of Physics, 48, 143

\bibitem[{{Taylor} {et~al.}(1999){Taylor}, {Carilli}, \& {Perley}}]{taylor99}
{Taylor}, G.~B., {Carilli}, C.~L., \& {Perley}, R.~A., eds. 1999, ASP Conf.
  Ser., Vol. 180, {Synthesis Imaging in Radio Astronomy II}

\bibitem[{{van Weeren} {et~al.}(2009){van Weeren}, {R{\"o}ttgering},
  {Br{\"u}ggen}, \& {Cohen}}]{vanweeren09}
{van Weeren}, R.~J., {R{\"o}ttgering}, H.~J.~A., {Br{\"u}ggen}, M., \& {Cohen},
  A. 2009, \aap, 508, 75

\bibitem[{{Vollmer} {et~al.}(2010){Vollmer}, {Gassmann}, {Derri{\`e}re},
  {Boch}, {Louys}, {Bonnarel}, {Dubois}, {Genova}, \& {Ochsenbein}}]{vollmer10}
{Vollmer}, B. {et~al.} 2010, \aap, 511, A53+

\end{thebibliography}
\end{document}